\documentclass[%
nofootinbib,
 aip,
 amsmath,amssymb,
 reprint,%
]{revtex4-1}

\usepackage{graphicx}
\usepackage{dcolumn}
\usepackage{bm}

\usepackage[utf8]{inputenc}
\usepackage[T1]{fontenc}

\usepackage{dsfont}
\usepackage{booktabs}
\usepackage[colorinlistoftodos]{todonotes}

\usepackage{xcolor}

\def\Q#1{\quad |q| = #1 \quad}

\begin{document}


\title[]{Sparsity-driven synchronization in oscillators networks}

\author{Antonio Mihara}
 \email{mihara@unifesp.br}
 \affiliation{Departamento de Física, Universidade Federal de São Paulo,UNIFESP, 09913-030, Campus Diadema, São Paulo, Brasil}
\author{Everton S. Medeiros}
 \email{medeiros@tu-berlin.de}
  \affiliation{Institut für Theoretische Physik, Technische Universität Berlin, Hardenbergstraße 36, 10623 Berlin, Germany}
\author{Anna Zakharova}
 \email{anna.zakharova@tu-berlin.de}
  \affiliation{Institut für Theoretische Physik, Technische Universität Berlin, Hardenbergstraße 36, 10623 Berlin, Germany}
  \author{Rene O. Medrano-T}
 \email{rene.medrano@unifesp.br}
 \affiliation{Departamento de Física, Universidade Federal de São Paulo,UNIFESP, 09913-030, Campus Diadema, São Paulo, Brasil}
\affiliation{Departamento de Física, Instituto de Geociências e Ciências Exatas,
Universidade Estadual Paulista, UNESP, 13506-900, Campus Rio Claro, São Paulo, Brasil}

\date{\today}

\begin{abstract}
 The emergence of synchronized behavior is a direct consequence of networking dynamical systems. Naturally, strict instances of this phenomenon, such as the states of complete synchronization are favored, or even ensured, in networks with a high density of connections. Conversely, in sparse networks, the system state-space is often shared by a variety of coexistent solutions. Consequently, the convergence to complete synchronized states is far from being certain. In this scenario, we report the surprising phenomenon in which completely synchronized states are made the sole attractor of sparse networks by removing network links, the {\it sparsity-driven synchronization}. This phenomenon is observed numerically for nonlocally coupled Kuramoto networks and verified analytically for locally coupled ones. In addition, we reduce the network equations to a one-dimension dynamical system to unravel the bifurcation scenario underlying the network transition to completely synchronized behavior. Furthermore, we present a simple procedure, based on the bifurcations in the thermodynamic limit, that determines the minimum number of links to be removed in order to ensure complete synchronization. Finally, we propose an application of the reported phenomenon as a control scheme to drive complete synchronization in high connectivity networks.
\end{abstract}

\maketitle

\begin{quotation}

Knowing the available dynamical states for networked systems is essential for planning and designing their coupling structure. This knowledge requires stability analysis of mathematical models capturing aspects of interest of the underlying real-world system. From this perspective, an open debate in the field concerns the connectivity constraints that would ensure the consensus dynamics among all the network units. Currently, the critical connectivity value for such a condition in any network of Kuramoto oscillators is $75\%$. For these systems, the consensus dynamics appear in form of a {\it {\bf in-phase synchronous state}} (ISS) which is globally stable for connectivities higher than the aforementioned critical value. This threshold is considered high for practical purposes. Fortunately, it has been also found in the literature that in sparse networks where usually different states coexist, one can globally stabilize the ISS by adding $O(n \log_2 n)$ links to the coupling structure. Here, we contribute to this field by introducing the phenomenon of sparsity-driven synchronization in regular networks of Kuramoto oscillators. This new phenomenon is counterintuitive once, contrary to the common sense of establishing synchronization by adding network links, it proposes the emergence of the ISS as the sole network attractor by removing network links in a coordinated way. The sparsity-driven synchronization occurs in locally and nonlocally coupled networks of Kuramoto oscillators with symmetric ring topology. The central strategy for the appearance of this phenomenon is a systematic removal
of links aiming at the eventual conversion of a closed ring topology into an open one.
\end{quotation}

\section{Introduction}
\label{sec:intro}

Often, the collective dynamics of systems ranging from the flashing of fireflies and the spiking of neurons to power grids and nanoelectromechanical oscillators exhibit synchronized behavior \cite{Winfree1967,Kuramoto1984, MirolloStrogatz1990,WatanabeStrogatz1994,BookSynchronization2001,Boccaletti2018,Science2019,Raphaldini2020}. The onset of different levels of synchronization is attributed to the interplay between the network structure and the local dynamics of each network component \cite{Jadbabaie2004,WSG2006,Taylor2011,Pikovsky2015,Poel2015,RPJK2016,Medeiros2019,Medeiros2021}. In particular, the so-called complete synchronized behavior is of interest to applications such as distributed systems of sensors \cite{YICK2008}, low-power radios \cite{Dokania2011}, and opinion formation \cite{Degroot1974,Baumann2020}.

From the theoretical point of view, many insights have been gathered from regular networks of identical Kuramoto oscillators. In such systems, the complete synchronized states correspond to {\it in-phase synchronous state} (ISS) occurring for sufficiently connected units. In the last two decades, systematic studies show that networks with high connectivity provide the suitable conditions for ISS to be globally stable, i.e. it is the sole network attractor \cite{WSG2006,Taylor2011,TSS2020}. Conversely, in sparse networks, the ISS is often sharing the system state-space with other synchronization patterns such as $q$-twisted states \cite{WSG2006,Canale2015,NossoNODY2019}, traveling waves \cite{Laing2016,Dou2018,NossoNODY2019}, solitary and chimera states \cite{Rybalova2019,Schuelen2020,Hellmann2020,Zakharova2020}. In such a scenario, the ISS possess a domain of attraction, i.e., a finite portion of the network state-space from where all trajectories converge to the ISS. This domain of attraction is called {\it sync basin}.  

Analyzing the sync basin, in $2006$, Wiley et al. \cite{WSG2006} observed that each oscillator in a regular network of identical Kuramoto oscillators must be connected to at least $\sim 68\%$ of its first neighbors in order to ensure global stability to ISS. In $2012$, Taylor \cite{Taylor2011} analyzed the necessary conditions for global ISS in networks with arbitrary structures ranging from regular to random. He obtained the minimal connectivity of $\sim 93\%$ for these cases. Following, additional studies have lowered this threshold to $\sim 79\%$ \cite{lu2020}. With this, comparing the minimal values for regular and arbitrary networks, the critical connectivity shall be between $\sim 68\%$ and $\sim 79\%$. In fact, recently, Townsend et al. \cite{TSS2020} conjectured that in any network of identical Kuramoto oscillators, such critical connectivity for the occurrence of globally stable ISS is $75\%$. For several networked systems in which the safe functioning depends on the ISS being globally stable, such required levels of connectivity can be impracticable. Fortunately, in the same aforementioned study, Townsend et al. \cite{TSS2020} have demonstrated that, even for sparse networks, the ISS can be globally stabilized by strategically adding $O(n \log_2 n)$ connections to the network. On one hand, this characteristic offers the possibility of beforehand designing the sparse network prescribed to exhibit global ISS. On the other hand, for systems already functioning, the real-time control by adding links is difficult due to practical issues of installing new physical connections in the network. 

Contrarily to these previous studies, we introduce the concept of sparsity-driven synchronization in oscillators networks. Auspiciously, this phenomenon does not require adding links to ensure complete synchronization in sparse networks. In fact, we demonstrate numerically and analytically that the systematic removal of links can globally stabilize the ISS in the state-space of regular networks of Kuramoto oscillators. First, we employ numerical simulations to show that all out-of-phase states coexisting with the ISS are systematically extinct for many realizations of locally and nonlocally coupled networks with symmetric ring topologies. We point out that the link removal strategy aims at the opening of the ring topology that may eventually result in a linear structure. Next, by performing linear stability analysis in locally coupled networks, we analytically verify the conversion of the ISS into the sole network attractor after the link removal procedure. Following these achievements, we obtain a low-dimensional equation for the network to unravel the bifurcation scenario providing global stability to the ISS. Furthermore, we develop a simple model of the network parameters able to predict the minimal number of links that must be removed from the ring for reaching global stabilization of the ISS.

This work is organized as follows: In section \ref{sec:sparsity_driven}, we first present the numerical evidence for the sparsity-driven synchronization phenomenon. We show numerical simulations for many different realizations of locally and nonlocally coupled Kuramoto oscillators (subsection \ref{sub:Rem_stab}). Next, we demonstrate the efficiency of the stabilization method as a control scheme to drive synchronization in the system (subsection \ref{sub:procedure_drive}). Following in section \ref{sec:sparsity_driven}, we perform a linear stability analysis to demonstrate the onset of a global ISS in a network with local coupling (subsection \ref{sub:linear_stability}). In section \ref{sec:how_sparsity_driven}, we focus on the mechanism in which sparsity-driven synchronization emerges. We perform a bifurcation analysis in a locally coupled network unveiling details of the transition to the globally stable ISS (subsection \ref{sub:bif_analysis}). Continuing in section \ref{sec:how_sparsity_driven}, for nonlocal networks, we study the minimum number of links to be removed in order to ensure global stability to the ISS. The simple model for this number is introduced (subsection \ref{sub:RemlinksVSattractors}). In section \ref{sec:conclusion}, we summarize our results. Simulations, plots and analysis were performed with Python and the libraries/packages Numpy, SciPy, Matplotlib, Networkx and JiTCODE \cite{Jitcode2018}.

\section{The onset of the globally stable in-phase synchronous state}
\label{sec:sparsity_driven}

In this section, we show numerical evidence, and an analytical confirmation, for the occurrence of the sparsity-driven synchronization in networks of Kuramoto oscillators. We also illustrate the applicability of this phenomenon as a control method based on the removal of network links.

\subsection{Sparse network model}
 \label{sub:Kum_rem}

We first define the dynamical and the topological aspects of the networks considered in throughout this work. Next, we investigate numerically the effects of removing networks links in a coordinated way aiming at the transformation from a ring to linear network topology.    

The dynamics of a network of identical Kuramoto oscillators is governed by the following set of equations:
\begin{equation}
    \dot{\theta_j} \equiv \frac{d\theta_j}{dt} = \omega + \sum_{k=1}^{N} G_{kj} \,\sin ( \theta_k - 
    \theta_j ) \,\, , j = 1, 2, ..., N \, .
\end{equation}
The phase of the $j$-th oscillator is $\theta_j$, and all oscillators have the same natural frequency $\omega$. The matrix $G$ is proportional to the adjacency matrix $A$: $ G_{kj} = \mathcal{G}(k,j) A_{kj} $, where $A_{kj}=A_{jk}$ is $1$ ($0$) if there is (not) a connection between oscillators $k$ and $j$. The function $\mathcal{G}(k,j)$ is the coupling strength between $k$ and $j$ which is assumed here to be symmetric and equally attractive: $\mathcal{G}(j,k) = \mathcal{G}(k,j) = g/(N-1) > 0$ ($N>1$). Additionally, one can change the variables to the rotating frame $\theta_j \rightarrow \theta_j + \omega t$, and further hide the (positive) constant $g/(N-1)$ by rescaling time, hence:
\begin{equation}
    \dot{\theta_j}  = 
    \sum_{k=1}^{N} A_{kj} \,\sin ( \theta_k - 
    \theta_j ) \, .
    \label{eq:Kuramoto}
\end{equation}

\begin{figure}[htb]
\includegraphics[width=7.5cm,height=3.5cm]{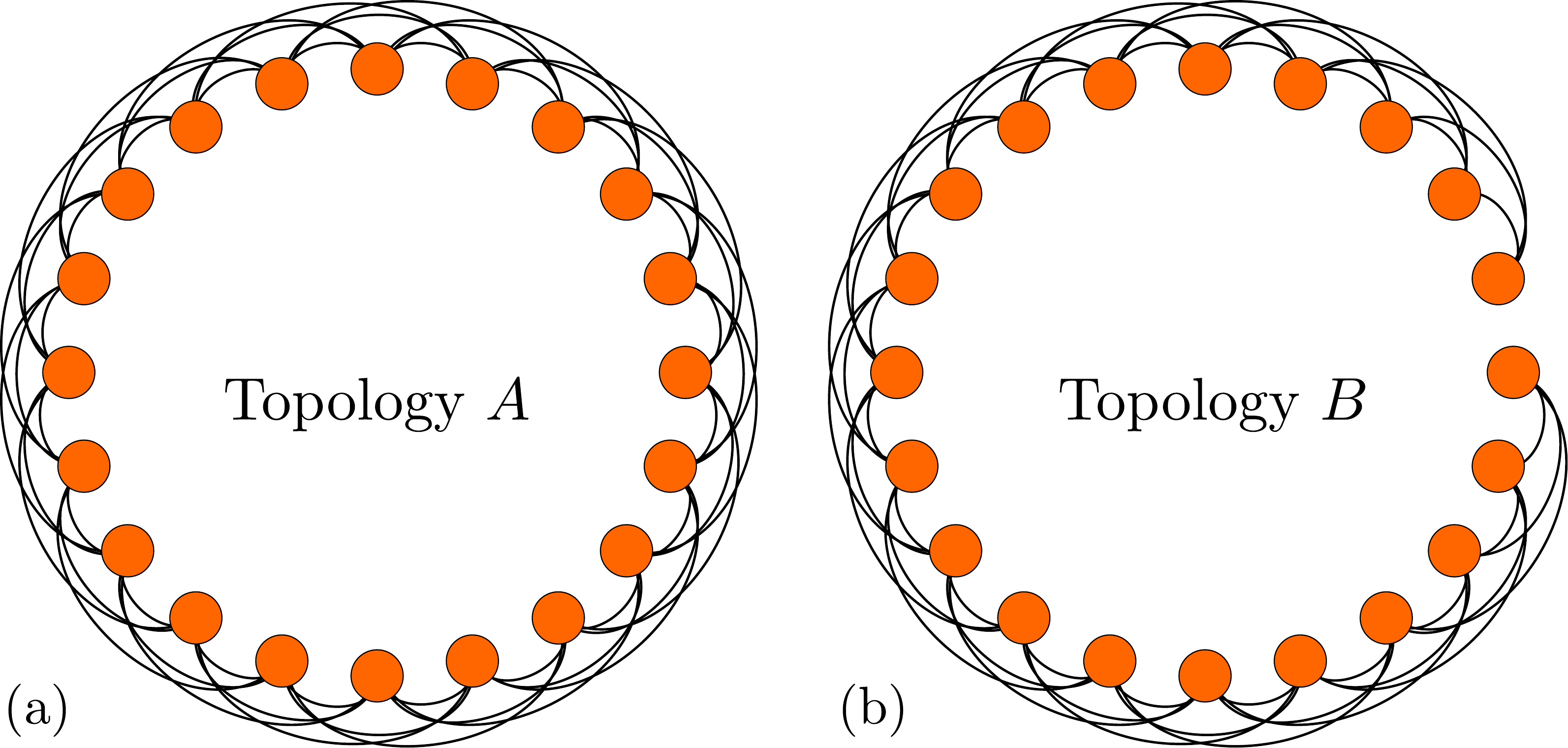}
\caption{Schematic of the nonlocally coupled network. (a) Regular ring topology with $N=20$ and $R=3$ (Topology $A$). (b) Open ring topology with  $\mathcal{L} = 6$ links removed from the regular ring (Topology $B$).}
\label{fig:method}
\end{figure}

In this study, we are mainly interested in regular networks connected with periodic boundary conditions forming a ring topology. As shown in Fig. \ref{fig:method}(a), each oscillator is coupled with equal strength to its $R$ nearest neighbors on either side. With this, Eq.~(\ref{eq:Kuramoto}) can be rewritten in a simpler form as: 
\begin{equation}
    \dot{\theta_j}  = 
    \sum_{k=j-R}^{j+R} \,\sin ( \theta_k - \theta_j ) \, ,
    \label{eq:WSG}
\end{equation}
where the index $k$ is periodic mod $N$.

The system described by Eq.~(\ref{eq:WSG}) has a number of equilibrium states (in the rotating frame) which can be characterized by:
\begin{equation}
    \theta_j = \frac{2\pi q}{N} j + C \, , 
\label{eq:solu}
\end{equation}
where $q$ is an integer called ``winding number'' specifying the number of full twists in the phase as a function of cycles on the ring. This quantity can assume the values $q=0,1,2,..., N-1$; and $C$ is an arbitrary constant. The state with $q=0$ corresponds to the ISS, i.e., $\theta_j = C, \forall j$. States with $q\neq 0$ are also known as ``$q$-twisted'' states.

Note that $q=N-1, N-2, N-3, ...$ is equivalent to $q=-1, -2, -3, ...$. In order to express this symmetry, from now we denote the twisted states as $q=\pm 1, \pm 2, \pm 3, ...$. The (linear) stability of such equilibria depends on the ratio $R/N$ ($\propto$ connectivity) : (i) for $R/N \gtrsim 0.34$, the network is considered dense, consequently, the ISS ($q=0$) is the only stable equilibrium; (ii) below this threshold, and as the ratio decreases, the network becomes sparse and more twisted states (with $q\neq 0$) become stable, reducing the size of sync basin. Hence, the sparse ring topologies such as the one shown in Fig.~\ref{fig:method}(a) with $N=20$ and $R=3$ ($R/N = 0.15$) possess multiple coexistent stable $q$-states, namely, $q = 0$ and $\pm 1\,$ \cite{NossoNODY2019}. 

\subsection{Removing links vs. stability}
 \label{sub:Rem_stab}

\begin{figure}[ht]
    \centering
    \includegraphics[width=0.48\textwidth]{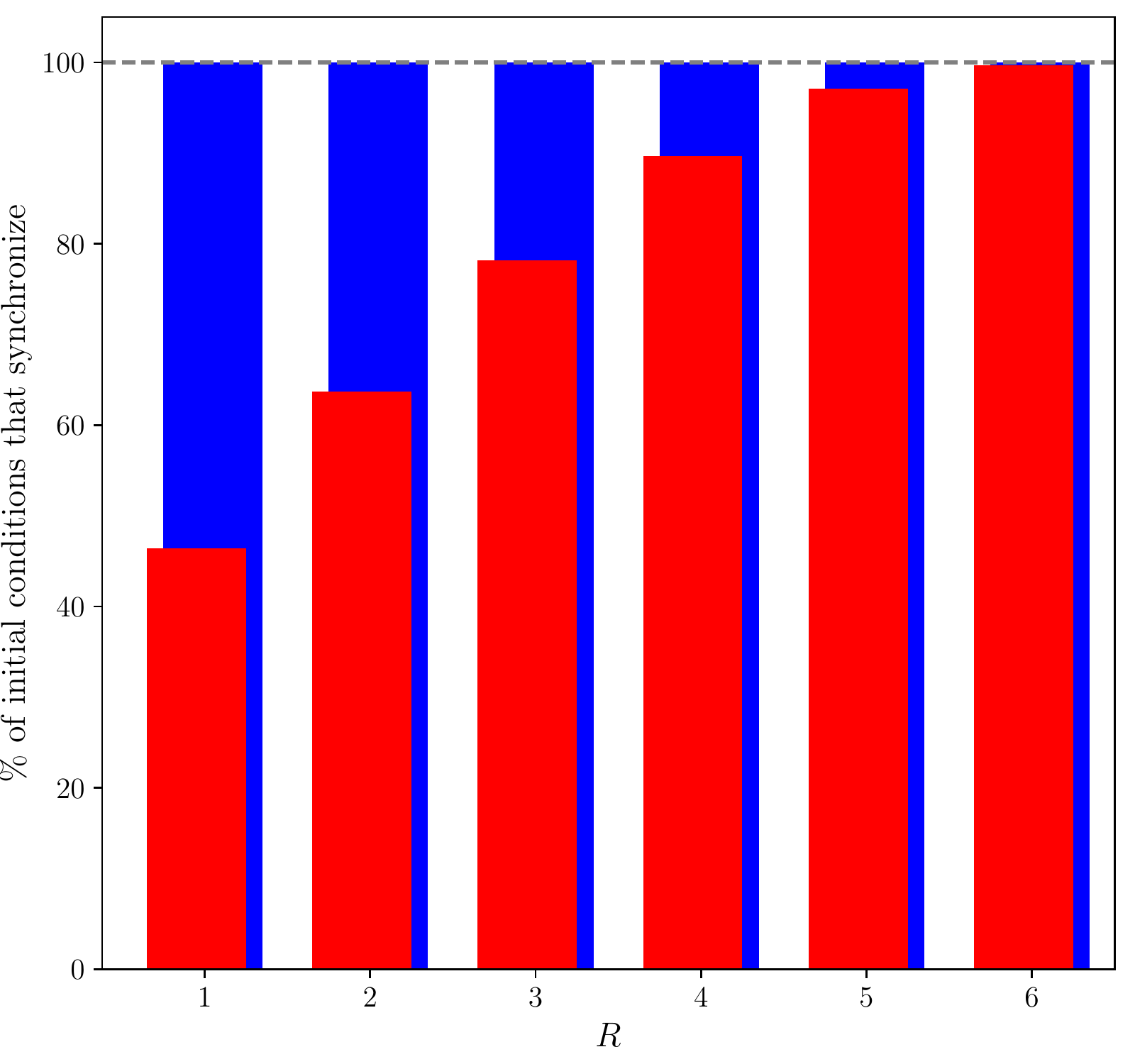}    
    \caption{Fraction of network realizations with different initial conditions leading to the in-phase synchronous state (ISS) for different values of $R$. The red color marks the results for the complete ring topology. The blue color shows the results for network realizations after the removal of $\mathcal{L}$ links (Eq. (\ref{eq:L})), the open ring topology. The network size is fixed at $N=20$. }
    \label{fig:fraction}
\end{figure}

 In order to promote the ISS as the sole stable equilibrium in such sparse networks, we propose a link removal strategy in which the ring topology may ultimately become open, as depicted in Fig.~\ref{fig:method}(b). We quantify the effectiveness of this strategy by numerically estimating the likelihood of the ISS to be reached for different network realizations with trajectories starting at random initial conditions. Hence, in Fig.~\ref{fig:fraction}, for $1000$ realizations of a network containing $N=20$ oscillators, the red bars shows the fraction of initial conditions (ICs) attributed to the network with the complete ring topology that converges to the ISS for different values of $R$. As expected, for low values of $R$ (sparse networks) the fraction of ICs converging to the ISS is low, indicating that the network state-space is shared among different $q$-states. In addition, this data agrees with previous results from the literature showing that the sync basin grows by making the network denser, i.e., increasing $R$ favors the convergence to the ISS. Now, we investigate the convergence characteristics of $1000$ realizations of the network with the open ring topology, i.e., after systematically removing a number $\mathcal{L}$ of links aiming at the opening of the ring topology.  As each node is coupled to its $R$ nearest neighbors on either side, one must remove a number of links given by the arithmetic series $\mathcal{L} = 1 + 2+...+R$ to achieve the open ring topology, so:
\begin{equation}
    \mathcal{L} = \frac{R}{2}(1+R).
\label{eq:L}
\end{equation}
With this, the fraction of network realizations leading to the ISS for the open topology is shown by the blue bars in Fig.~\ref{fig:fraction}. Interestingly, after such link removal, the ISS is reached for $100\%$ of the realizations for all values of $R$.    

Hence, the data shown in Fig. \ref{fig:fraction} provides numerical evidence for the ISS to be the sole network attractor after the removal
of $\mathcal{L}$ links aiming at open ring topologies. We point out that the removal of all $\mathcal{L}$ is not a necessary condition to make the ISS globally stable. Depending on the network parameters $N$ and $R$, the same stability gain can be obtained by removing a minimal number $\mathcal{L}_m$ of links. This fact, together with analytical evidence for our claiming as well as the bifurcation scenario underlying the process, will be discussed in the upcoming sections. In the next section, we demonstrate the applicability of the link removal procedure to drive complete synchronization in the network of Kuramoto oscillators.

\subsection{Driving sparse networks into synchronization}
\label{sub:procedure_drive}

\begin{figure}[htp]
\centering
\includegraphics[width=0.48\textwidth]{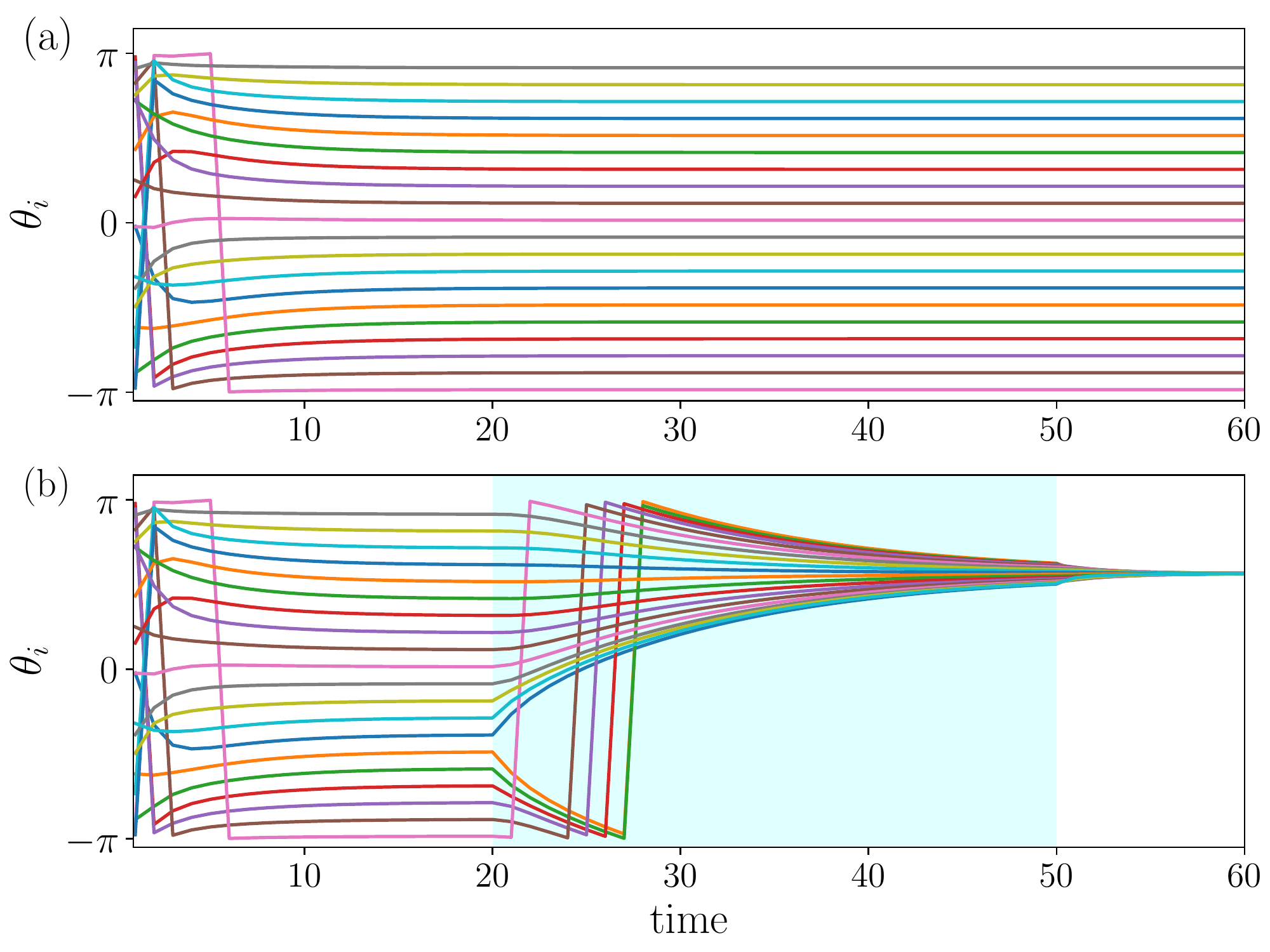}
\caption{Time evolution of the oscillators for two simulations obtained from the same initial condition. (a) For topology $A$, the network is numerically integrated reaching the state $q=-1$. (b) The network is initialized in topology $A$ (see Fig. \ref{fig:method}(a)). At $t=20$, the structure is temporarily switched to topology $B$ (see Fig. \ref{fig:method}(b)) up to $t=50$ at which topology $A$ is restored. $N=20$ and $R=3$.}
\label{fig:resN20R3} 
\end{figure}

As shown in the previous section, the coordinated removal of network links is able to destabilize $q$-states competing with the ISS for stability in sparse networks. Since the recent literature mainly supports the onset of completely synchronized behavior in sparse networks via the addition of extra network links \cite{TSS2020}, the counterintuitive procedure reported here is innovative. It may have significant applicability especially because the eventual disconnection of network links is intrinsically a more practical procedure than the installation of new ones. To illustrate such applicability, we initially consider a network composed of $N=20$ oscillators nonlocally coupled in the ring topology with $R=3$ on each side, see topology $A$ in Fig. \ref{fig:method}. For this topology, the twisted $q$-states with $q= \pm 1$ are stable (see Fig. 3(a) of Ref. \cite{NossoNODY2019}). In Fig. \ref{fig:resN20R3}(a), we show the time evolution of the network in this state. In this simulation the network trajectories are numerically integrated up to $t = 60$, the system does not fully synchronize reaching the twisted state $q=-1$. Notice the out-phase behavior of each oscillator. Now, in order to demonstrate the effectiveness of considering a less dense network to stimulate in-phase behavior, we intervene in the system dynamics by considering the open ring, topology $B$ in Fig. \ref{fig:method}, for a finite interval of time. Subsequently, the topology $A$ is restored forming the A-B-A procedure. More specifically, in Fig. \ref{fig:resN20R3}(b), for $t\leq 20$ the network trajectories are obtained for the system in the topology $A$ for the same initial conditions adopted in in Fig. \ref{fig:resN20R3}(a). Then, for $20 < t \leq 50$, the network is set in the topology $B$ with the removal of $\mathcal{L} = 6$ links (cyan shaded area of Fig. \ref{fig:resN20R3}(b)). During this time interval, the network trajectories approach the ISS. Finally, for $t > 50$, the topology $A$ is restored without affecting the fully synchronized behavior achieved with the topology $B$.

\subsection{Linear stability analysis: the conversion of the ISS into the sole network attractor}
\label{sub:linear_stability}

\begin{figure}[htp]
\includegraphics[width=7.5cm,height=3.5cm]{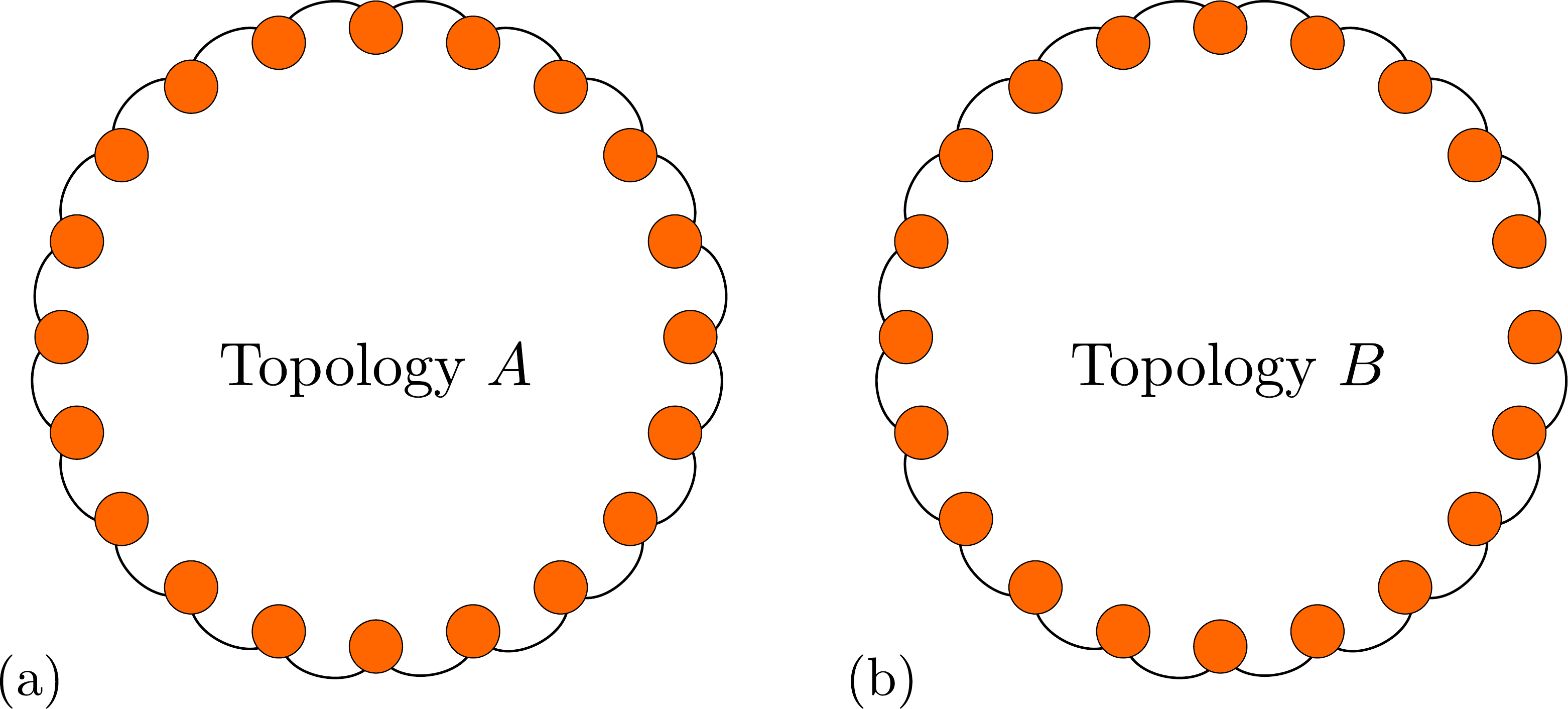}
\caption{Schematic of the locally coupled network used for the linear stability analysis. (a) Regular ring network with $N=20$ and $R=1$ (Topology $A$). (b) The open ring network with one link removed from the regular ring (Topology $B$).}
\label{fig:local}
\end{figure}

The onset of the ISS as a globally stable attractor of the network of Kuramoto oscillators has been numerically demonstrated in this section for networks with different link densities. These results are shown in Fig. \ref{fig:fraction} where one can see that the lowest fraction of network realizations converging to the ISS occurs for the complete ring with $R=1$, i.e., the most sparse ring network (see topology $A$ in Fig. \ref{fig:local}). With this, we choose this particular case to perform a linear stability analysis and demonstrate that the ISS would be the sole stable network attractor after the removal of $\mathcal{L}=1$ links resulting in an open ring as topology $B$ in Fig. \ref{fig:local}. 

The dynamical equation describing the network of Kuramoto oscillators in a complete ring (topology $A$ Fig. \ref{fig:local})) is given by:
\begin{equation}
\begin{cases}
\dot{\theta_1} &= \sin( \theta_N - \theta_1 ) 
+ \sin( \theta_2 - \theta_1 )  \\
\dot{\theta_2} &= \sin( \theta_1 - \theta_2 ) 
+ \sin( \theta_3 - \theta_2 )  \\
&\vdots \\
\dot{\theta_N} &= \sin( \theta_{N-1} - \theta_N ) 
+ \sin( \theta_1 - \theta_N )  \, .
\end{cases}
\label{eq:sysA0}
\end{equation}

Now, in order to simplify the notation, we define the variables $x_j \equiv \theta_{j+1} - \theta_j$, $j = 1, 2, ..., (N-1)$ and 
\begin{equation}
    z \equiv \sum_{j} x_j = \theta_N - \theta_1 \,.
\end{equation}
With this, Eq. (\ref{eq:sysA0}) can be rewritten as:
\begin{equation}
\begin{cases}
\dot{x}_1 &= - \sin(z) - 2\sin( x_1 ) + \sin( x_2 )  \\
\dot{x}_2 &= \sin(x_1) - 2\sin(x_2 ) + \sin( x_3 )  \\
&\vdots \\
\dot{x}_{N-2} &=  \sin(x_{N-3}) - 2\sin( x_{N-2} ) + \sin( x_{N-1} )\\
\dot{x}_{N-1} &= - \sin(z) - 2\sin( x_{N-1} ) + \sin( x_{N-2} ) \, .
\end{cases}
\label{eq:sysA}
\end{equation}

Once the solution of the network described in Eq. (\ref{eq:sysA0}) is given by Eq. (\ref{eq:solu}), the solution of the system in Eq. (\ref{eq:sysA}) is given by $x_j^* = \delta = (2\pi q)/N$, $j=1,2,..., (N-1)$, with $z^* =-\delta$. With this, the eigenvalues of the system Jacobian matrix evaluated at $x_j^*$ are given by \cite{NossoNODY2019}:
\begin{equation}
    \gamma_{\ell} = -4 \cos\left( \frac{2\pi q}{N} \right)
    \sin^2 \left( \frac{\pi \ell}{N} \right) \, ,
    \quad \ell = 1, ..., (N-1).
    \label{Eq:eigenvalues_complete}
\end{equation}
The condition for stability is $\gamma_{\ell} < 0 $, which implies that $|q|<N/4$. Hence, the ISS ($q=0$) is always a stable state, however, a collection of twisted states ($q\neq 0$) are also stable for $N \geq 5$. 

Next, we now investigate the linear stability of the network after the removal of $\mathcal{L}=1$ link (topology $B$ in Fig. \ref{fig:local})). We remove the link connecting the oscillators $1$ and $N$, the dynamical equation of such an open ring network is given by:
\begin{equation}
\begin{cases}
\dot{x}_1 &= - 2\sin( x_1 ) + \sin( x_2 )  \\
\dot{x}_2 &= \sin(x_1) - 2\sin(x_2 ) + \sin( x_3 )  \\
          &\vdots \\
\dot{x}_{N-2} &=  \sin(x_{N-3}) - 2\sin( x_{N-2} ) + \sin( x_{N-1} )\\
\dot{x}_{N-1} &= - 2\sin( x_{N-1} ) + \sin( x_{N-2} ) \, .
\end{cases}
\label{eq:sysB}
\end{equation}

First, we assume the solution of Eq. (\ref{eq:sysB}) in the form $x_j = \delta =$ constant $\forall j$. Replacing this solution into Eq. (\ref{eq:sysB}), we obtain:
\begin{equation}
\begin{cases}
0 &= - 2\sin( \delta ) + \sin( \delta )  \\
0 &= \sin(\delta) - 2\sin(\delta ) + \sin( \delta )  \\
  &\vdots \\
0 &= \sin(\delta) - 2\sin(\delta ) + \sin( \delta )  \\
0 &= - 2\sin( \delta ) + \sin( \delta ) \, .
\label{eq:sysB2}
\end{cases}
\end{equation}
Hence, we find $\sin(\delta) = 0$. Therefore, the solution is $x_j ^* = \delta = m\pi$, $m \in \mathbb{Z}$. The corresponding Jacobian matrix evaluated at this solution is a tridiagonal Toeplitz matrix:
\begin{equation}
J = 
\begin{pmatrix}
a & b & 0 & \cdots & \cdots & \cdots & \cdots & 0\\
c & a & b & 0 & & & & \vdots\\
0 & c & a & b & \ddots & & & \vdots\\
\vdots & 0 & \ddots & \ddots & \ddots & \ddots & & \vdots\\
\vdots & & \ddots & \ddots & \ddots & \ddots & 0 & \vdots\\
\vdots & & & \ddots & c & a & b & 0\\
\vdots & & & & 0 & c & a & b\\
0 & \cdots & \cdots  & \cdots & \cdots & 0 & c & a\\
\end{pmatrix}
\end{equation}
where $a = -2 \cos(m\pi) = (-2)(-1)^m$ and $b = c = \cos(m\pi) = (-1)^m$. The eigenvalues are given by:
\begin{equation}
    \gamma_{\ell} = -2 \, \left[ (-1)^m + 
    \cos \left( \frac{\pi \ell}{N} \right) \right] \, ,
    \; \ell = 1, ..., (N-1).
    \label{Eq:eigenvalues_open}
\end{equation}
From Eq. (\ref{Eq:eigenvalues_open}), we conclude that stable solution for the network after the removal of $\mathcal{L}=1$ link occurs  only for $m$ even, i.e., $x_j ^* = 0$ (remind that the variables $x_j$ are mod $2\pi$). Therefore, this result, associated with the blue bar ($R=1$) in Fig. \ref{fig:fraction}, indicates that the sole stable attractor in the network with open ring topology is the one with $q=0$, i.e., the in-phase synchronous state (ISS).


\section{How sparsity drives Kuramoto networks to synchronization}
\label{sec:how_sparsity_driven}

So far, we have demonstrated numerically for nonlocal networks, and analytically for local ones, that the removal of network links aiming at the transformation from a closed to an open ring topology makes the ISS globally stable. In this section, we address the mechanisms underlying such a process.

\subsection{Bifurcation analysis for local networks}
\label{sub:bif_analysis}

Our main goal in this subsection is to reveal how all the $q$-twists states lose stability making the $q=0$ (ISS) the sole attractor of Kuramoto networks. For that, in a local ring topology ($R=1$) composed of $N$ oscillators, we attribute a weight $w$ to the intensity of one of the network links:
\begin{equation}
\begin{cases}
\dot{x}_1 &= - w \sin(z) - 2\sin( x_1 ) + \sin( x_2 )  \\
\dot{x}_2 &= \sin(x_1) - 2\sin(x_2 ) + \sin( x_3 )  \\
&\vdots \\
\dot{x}_{N-2} &=  \sin(x_{N-3}) - 2\sin( x_{N-2} ) + \sin( x_{N-1} )\\
\dot{x}_{N-1} &= - w \sin(z) - 2\sin( x_{N-1} ) + \sin( x_{N-2} ) \, .
\end{cases}
\label{eq:Simplif.Sys}
\end{equation}
This new parameter is defined in a continuous interval $w \in [0,1]$ such that $w=1$ rescues the complete ring topology (Eq. (\ref{eq:sysA})), where symmetric $q$-twisted states ($x_j=\delta$, $|z|=|\delta|$, and $\delta=2\pi q/N$) are stable, while $w=0$ opens the ring (Eq. (\ref{eq:sysB})), where ISS ($x_j=0$)  is the only stable state. Any other value of $w \in (0,1)$ leads the system to assymmetric equilibriums ($x_j=\delta$, $|z|\neq|\delta|$, and $\delta =$ constant $\neq 2\pi q/N$).

\begin{figure}[htp]
    \centering
    \includegraphics[width=8.5cm,height=4.5cm]{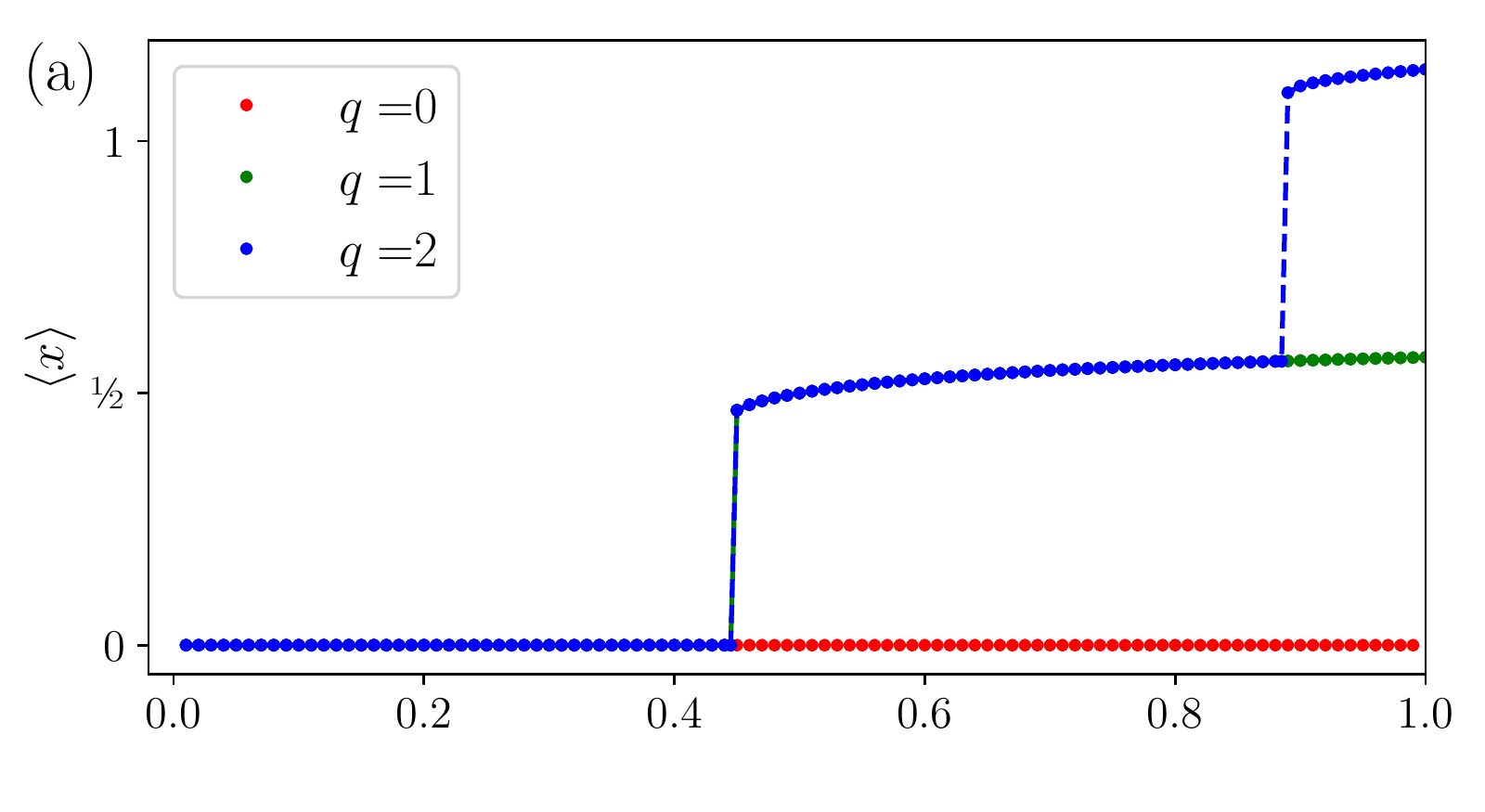}
    \includegraphics[width=8.5cm,height=4.5cm]{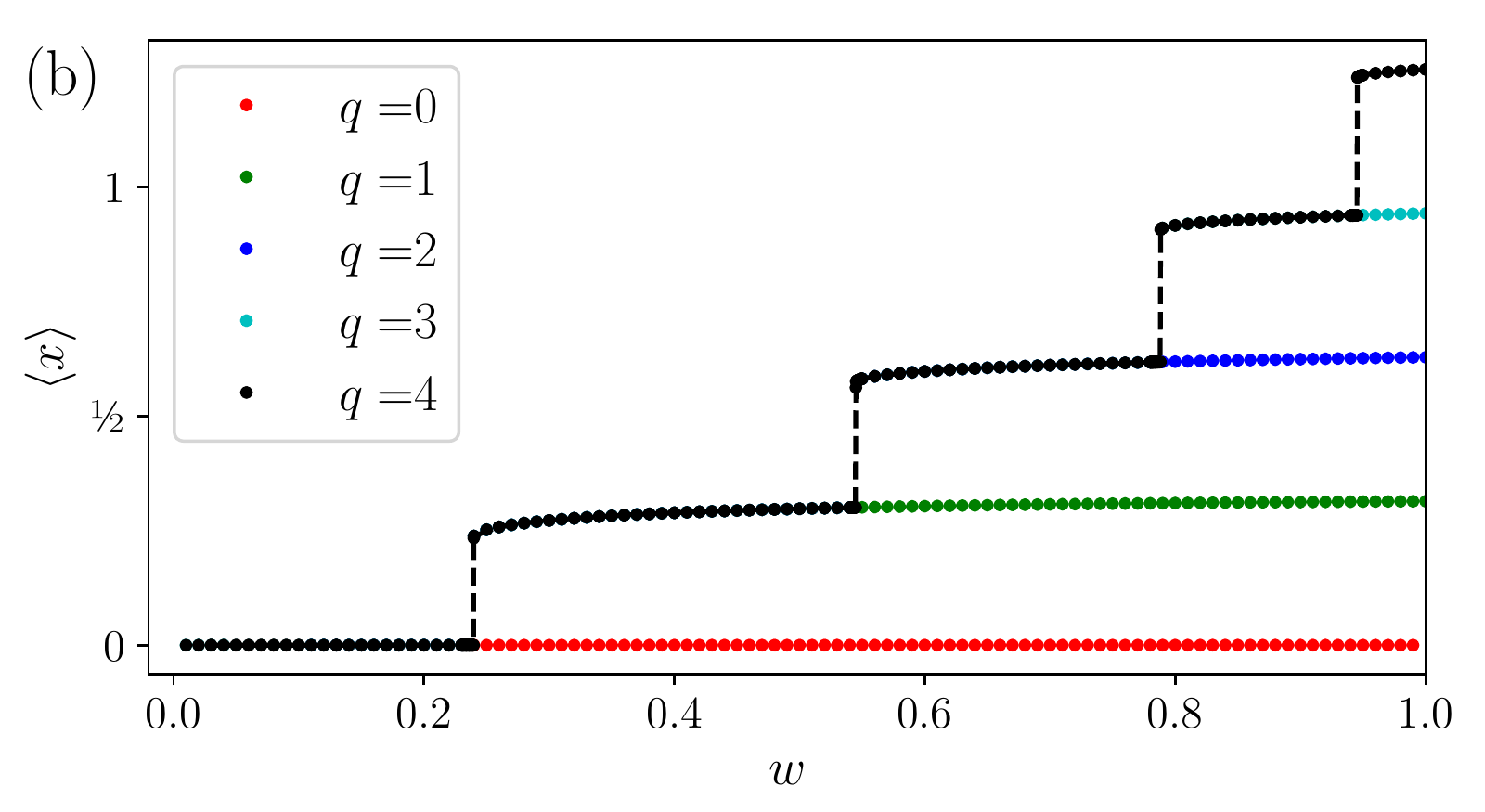}
    \caption{Bifurcation diagram for the average variable $\langle x \rangle$ obtained by following the attractors as the weight in the link intensity varies in the interval $w \in [0,1]$ in a locally coupled network ($R=1$). The configuration with $w=1$ corresponds to the complete ring topology, while $w=0$ corresponds to the removal of one network link. The color code stands for the different $q$-twisted states. (a) Network size: $N=11$. (b) Network size: $N=20$. The blue color drawing a ladder means that the state $q=2$ bifurcates to $q=1$ and then to $q=0$ from $\omega=1$ to 0. For a continuous increment of $w$, $\langle x \rangle$ does not change its $q$-twist state. The black ladder is equivalent.}
    \label{fig:escadaN11}
\end{figure}

\begin{figure*}[htp]
\centering
\includegraphics[width=0.95\textwidth]{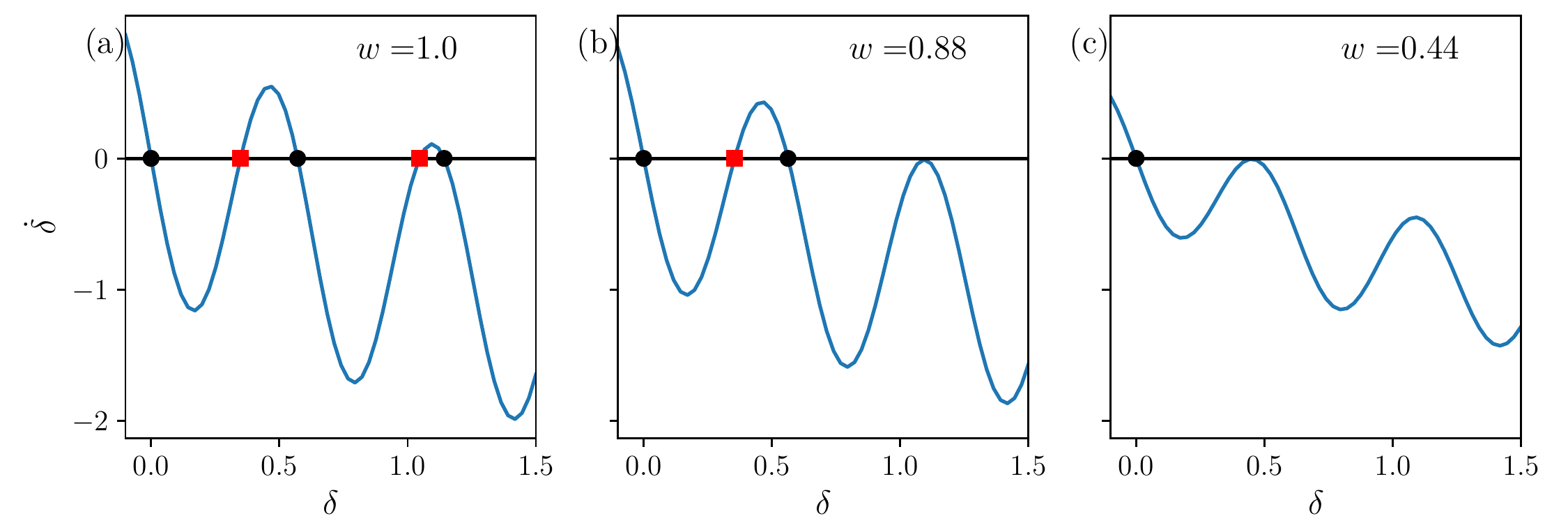}
\caption{State-space $(\delta, \dot{\delta})$ of the one-dimensional system shown in Eq. (\ref{eq:delta}). The black circles correspond to stable equilibria, while the red squares correspond to saddle points. The network size is fixed at $N=11$. (a) For $w=1$, the states $q=0$, $q=1$, and $q=2$ are stable. (b) For $w=0.88$, at this value, the saddle-node bifurcation destroys the state with $q=2$. (c) For $w=0.44$, at this value, the saddle-node bifurcation destroys the state with $q=1$.}
\label{fig:bifsN11} 
\end{figure*}

In order to gather insights on such changes, we obtain a bifurcation diagram for the average $\langle x \rangle = \sum_j x_j / (N-1)$ as the parameter $w$ varies in the interval $0<w<1$.
It is worth to mention that $x_j$ has standard deviation
of about $O(10^{-10})$ from $\langle x \rangle$ after the evolution transient time  $\Delta t = 500$ is 
considered, what is in agreement with the asymmetric equilibria $x_j = \delta$ with $\delta =$ constant $\neq 2\pi q/N$.
With this, in Fig. \ref{fig:escadaN11}(a), we show the results for the local network with $N=11$ oscillators. In this diagram, for $w=1$ the system has initially $5$ attractors, $q=0, \pm 1, \pm 2$. As $w$ decreases, states with large $q$ are affected first, hence, the state with $q=\pm 2$ undergoes a bifurcation first at $w \approx 0.88$ and the state with $q=\pm 1$ at $w \approx 0.44$. In Fig. \ref{fig:escadaN11}(b), we show the bifurcation diagram of a larger network $N=20$. For this case, despite having a larger number of stable states at $w=1$, the bifurcation scenario is very much alike to the one in Fig.  \ref{fig:escadaN11}(a). We point out that, for $w \lesssim 0.44$ in Fig. \ref{fig:escadaN11}(a) and $w \lesssim 0.23$ in Fig. \ref{fig:escadaN11}(b), the state with $q=0$ (ISS) is already the sole network attractor, i.e., the ISS becomes globally stable before the ring topology opens at $w=0$. This fact will be useful to investigate the minimum number $\mathcal{L}_m$ of links to be removed in order to destabilize all the $q$ states in a nonlocal network ($R>1$). Before we address this issue in the next section, we now unveil which bifurcation is mediating the transitions undergone by the $q$-twisted states observed in Fig. \ref{fig:escadaN11}. 
For this task, we perform a dimension reduction of the equations assuming the solutions of Eq. (\ref{eq:Simplif.Sys}) to be in the form $x_j = \delta$ and making $z=\sum_j x_j$ (Note that if $w=1$, $z=-\delta=-2\pi q/N$, as resulted in Eq. (\ref{eq:sysA})). With this, Eq. (\ref{eq:Simplif.Sys}) becomes:
\begin{equation}
\begin{cases}
\dot{\delta} &= - w \sin\left((N-1)\delta\right) - \sin( \delta )  \\
\dot{\delta} &= 0  \\
&\vdots \\
\dot{\delta} &= 0 \\
\dot{\delta} &= - w \sin\left((N-1)\delta\right) - \sin( \delta ) \, .
\end{cases}
\label{eq:Simplif.Sys. delta}
\end{equation}
Consequently, under these considerations, the system dynamics can be simply described by:
\begin{equation}
   \dot{\delta} = - w \sin\left((N-1)\delta\right) - \sin( \delta ).
   \label{eq:delta} 
\end{equation}
Although the roots of Eq. (\ref{eq:delta}) provide us the system equilibria as a function of the parameter $w$, the stability of each equilibrium must still be obtained via the eigenvalues of the Jacobian matrix of Eq. (\ref{eq:Simplif.Sys}). Hence, in Fig. \ref{fig:bifsN11}, we plot the curve given by Eq. (\ref{eq:delta}) for the network with $N=11$ oscillators (same as in Fig. \ref{fig:escadaN11}(a)). In this figure, the equilibria ($\dot{\delta}=0$) are marked in accordance with their respective stability, i.e., attractors are represented by black circles and saddles by red squares. For $w=1.0$ corresponding to the complete ring topology, in Fig. \ref{fig:bifsN11}(a) we observe the attractors with $q=0$, $q=1$, and $q=2$ with saddle points in their basin boundaries. As $w$ decreases, the attractors with $q=2$ and $q=1$ approach the saddle points in their respective basin boundaries. At $w=0.88$ shown in Fig. \ref{fig:bifsN11}(b), we verify that the attractor with $q=2$ is destroyed in a collision with a saddle point. This mechanism characterizes a saddle-node bifurcation. Subsequently, at $w=0.44$ shown in Fig. \ref{fig:bifsN11}(c), we observe the collision of the attractor with $q=1$ with another saddle point in a second saddle-node bifurcation. As a result, for $w < 0.44$ only the ISS (with $\delta=0$) remains stable.

In this subsection, we have addressed the bifurcation scenario responsible for the global stability of the ISS after removing one link of a locally coupled network. The introduction of weight to the intensity of one network link has allowed the analysis not only for the closed and open ring topology but also of the bifurcations in-between. One interesting finding is the ISS becoming globally stable even when the network link still possesses a significant intensity ($w \approx 0.44$). This fact suggests that the ISS can also become globally stable in a nonlocal network ($R>1$) without the necessity of removing $\mathcal{L}$ links. This is the subject of the next subsection.  

\subsection{Number of removed links vs. attractors in nonlocal networks}
\label{sub:RemlinksVSattractors}

Motivated by the results obtained for local networks in which the ISS become globally stable for $w \to 0$, in this subsection, we investigate the transformations undergone in the state-space of a nonlocal network under the systematic removal of links. More specifically, given the network parameters $N$ and $R>1$, we address the question of what would be the minimum number of links $\mathcal{L}_m$ that must be removed in order to globally
synchronize the network?

To tackle this issue, we first perform numerical simulations of a network with $N=40$ and $R=3$. For these parameters, the network in the complete ring topology possesses seven attractors in its state-space, namely: $q=0, \pm 1, \pm 2, \pm 3$. Next, we consider $1000$ network realizations with different initial conditions (ICs) to analyze the fraction of trajectories reaching each attractor for different numbers \# of removed links. The results are compiled in Table \ref{tab:ResultsN40R3}. In the first row of this table, we show the fraction of realizations converging to the attractors of the complete ring topology ($\#=0$). For this case, we observe that the state with $q=0$ is the most visited, indicating that it has the largest basin of attraction. On the other end, the state with $q=\pm 3$ is not visited, a consequence of a very small basin of attraction. 

Following these results for the complete ring topology, we now analyze the fraction of initial conditions converging to each attractor after the link removal procedure. The order of link removal obeys a distance rule, from the farthest to the closest. We start each network simulation with a number \# of removed links. The corresponding system is evolved for a time interval of duration $\Delta t=500$. At this point, the complete ring topology is restored, and the system is evolved for an extra time interval $\Delta t=500$. At the end of the time evolution, the final network state of each realization is recorded, and the respective fraction of convergence is shown in Table \ref{tab:ResultsN40R3}. The number of removed links necessary to open the ring topology for this network parameters is $\mathcal{L}=6$. However, in Table \ref{tab:ResultsN40R3}, we observe that for $\#=3$ the attractors with $|q|=2$ are already not visited for network trajectories. Furthermore, for $\#=5$, the attractors with $q=0$ (ISS) are the sole survivors in the system state-space. This is a numerical evidence  that the minimum number of links to be removed to ISS reaching global stability is $\mathcal{L}_m=5$. In this scenario, our next step is to further investigate and possibly generalize these observations.          

    \begin{table}[th]
        \centering
        \begin{tabular}{c|ccc}
    & \multicolumn{3}{c}{\% of initial conditions that} \\
$\#$ removed & \multicolumn{3}{c}{converged to:}\\
            links &  $\quad q=0\quad$ & $\Q1$ & $\Q2$ \\
            \midrule
            0 & 58.5 & 40.5 & 1.0  \\            
            1 & 58.0 & 41.1 & 0.9  \\ 
            2 & 58.4 & 40.3 & 1.3  \\ 
            3 & 59.2 & 40.8 & 0    \\
            4 & 60.8 & 39.2 & 0    \\
            5 & 100  & 0    & 0    \\
            6 & 100  & 0    & 0    \\
            \bottomrule
        \end{tabular}
        \caption{Results of simulations for $N=40, R=3$.}
        \label{tab:ResultsN40R3}        
    \end{table}

Differently from the network locally connected, an analytical approach to obtain $\mathcal{L}_m$ in a network with $R>1$ would require the analysis of complicated systems of ODEs. Therefore, to gather further insights on how sparsity (link removal) drives the Kuramoto network into completely synchronized behavior, we develop a simple model able to capture the essential behavior observed numerically. The goal is to obtain an approximation of the minimum number of removed links $\mathcal{L}_m$ for which the ISS reaches global stability in networks with arbitrary values of $R$ and $N$. For this, we first look at local networks analyzed in (the previous)
subsection \ref{sub:bif_analysis}. For that case, in a network containing $N_A$ attractors with $q\geq 0$ in the complete ring topology, but with a link with weight $w$, the bifurcation diagram for $\langle x \rangle$ as $w$ varies from 1 to 0 exhibits a ladder shape with $N_A-1$ steps as shows Fig. \ref{fig:escadaN11}. For our purposes, we assume that $N_A \gg 1$ and 
all steps are of the same size and, therefore, the critical weight below which the ISS ($\langle x \rangle = 0$) becomes globally stable would be estimated as $w_c \approx 1/N_A$. 

Now let us establish a correspondence between the model analyzed in subsection \ref{sub:bif_analysis} (a ring with local couplings) and the
procedure of link removal in ring with nonlocal couplings. Then we must relate the variables $w_c$, $N_A$ with $\mathcal{L}$, $\mathcal{L}_m$. For such an aim,  we call the attention that for nonlocal networks, the number of removed links varying in the interval $0 \leq \# \leq \mathcal{L}$  plays a similar role as $w$ in the interval $0 \leq w \leq 1$ for a local ring (as detailed in subsection \ref{sub:bif_analysis})  in the sense that removing links in nonlocal networks corresponds to diminishing the weight $w$. 
 By assuming a (naive) linear relation between these two scales, $\mathcal{L}_m$ (the minimum number of links required for ISS to become globally stable in nonlocal networks) and $w_c$ (the critical $w$ for the ISS bifurcation, estimated as $1/N_A$) are related by:

\begin{equation}
    \mathcal{L}_m \approx \mathcal{L}\,\left( 1 - \frac{1}{N_A}\right)\, .
    \label{eq:Lmin0}
  \end{equation}

Now, we use the fact that the attractors of networks of identical Kuramoto oscillators can be lebeled as $-q_{max}, \dots,-2,-1,0,1,2, \dots,q_{max}$ to express the number of attractors as $N_A = q_{max} + 1$, whereas $q_{max}$ is the largest winding number present in the network. Considering this $N_A$  and  $\mathcal{L}$, according Eq. (\ref{eq:L}), Eq. (\ref{eq:Lmin0}) can be written as:
\begin{equation}
    \mathcal{L}_{m} =  \left\lceil 
    \frac{R(1+R)q_{max}}{2(q_{max} + 1)} \right\rceil,
    \label{eq:Lmin1}
\end{equation}
where the operation $\lceil . \rceil$ stands for the ceil function defined as $\lceil x\rceil \equiv \min\{ n\in \mathds{Z} | n \geq x\}$, for $x\in \mathds{R}$. Fortunately, in the literature there are practical ways to evaluate $q_{max}$ for Kuramoto networks with arbitrary parameters $N$ and $R$ \cite{WSG2006,NossoNODY2019,Maistrenko2012} based on the connectivity fraction $r = (2R+1)/2N$ ($r=0.5$ indicates all-to-all coupling, i.e., $100\%$ of connectivity). The largest winding number $q_{max}$ is obtained by identifying the respective value of $r$ in the exact bifurcation scenario of a Kuramoto network in the thermodynamic limit. More specifically, Table \ref{tab:q_and_r} depicts results from the literature showing $q$ and its respective bifurcation point $r_c$ obtained for $N \rightarrow \infty$. This table should be read as for $0 < r \leq 0.5$ the ISS is stable, for $0 < r < 0.340461$ the twist state $|q|=1$ is stable, for $0 < r < 1/6$ the twist state $|q|=2$ is stable, and so on. The value of $q_{max}$ is given by the value of $q$ corresponding to the nearest $r_c \geq r$ shown in Table \ref{tab:q_and_r}. More details about building this bifurcation table is given in Ref. \cite{NossoNODY2019}.

\begin{table}[hbt]
 \centering
 \begin{tabular}{ccc}
  \toprule
  $|q|$ & $\quad$ & $r_c$   \\ 
  \midrule
    1 & & 0.340461  \\
    2 & & $1/6$   \\
3 & & 0.110727 \\
4 & & 0.082947 \\ 
5 & & 0.066322 \\
6 & & 0.055252 \\
7 & & 0.047351 \\
8 & & 0.041427 \\
9 & & 0.036821 \\
10 & &  0.033137 \\
11 & & 0.030123 \\
12 & & 0.027612 \\
\bottomrule
\end{tabular}
\caption{Bifurcation of $q$-twisted states according the connectivety fraction $r$ in the thermodinamic limit ($N \to \infty$).}
\label{tab:q_and_r}
\end{table}

To illustrate the use of Table \ref{tab:q_and_r} for a nonlocal network with $R=3$, we consider three different network sizes: i) For $N=21$, we compute $r=7/42=1/6$ and, from Table \ref{tab:q_and_r}, we obtain $q_{max}=2$. ii) For $N=40$, the quantity $r$ is $r=7/80=0.0875$. This value is between $0.082947$ and $0.110727$ in Table \ref{tab:q_and_r}, then we obtain $q_{max} = 3$. iii) For $N=80$, we have $r=7/160=0.04375$ and, therefore, $q_{max}=7$. Since that, for $R=3$ the number of removed links to completely open the ring is $\mathcal{L}=6$, Eq. (\ref{eq:Lmin1}) is $\mathcal{L}_{m} = \lceil6\, q_{max}/(q_{max}+1)\rceil$. Now, we substitute the respective values of $q_{max}$ to estimate the minimum number of links $\mathcal{L}_m$ for the global stability of the ISS. Hence, for the three aforementioned network sizes, we obtain:   
\begin{eqnarray*}
i) &\quad& N= 21, \quad q_{max}= 2, \quad  \mathcal{L}_{m} = 4 \nonumber\\
ii) &\quad& N= 40, \quad q_{max}= 3, \quad  \mathcal{L}_{m} = 5 \nonumber\\
iii) &\quad& N= 80, \quad q_{max}= 7, \quad  \mathcal{L}_{m} = 6.
\end{eqnarray*}
These estimates for $\mathcal{L}_{m}$ are all in agreement with numerical results including the introductory case shown in Table \ref{tab:ResultsN40R3}.

\begin{figure}[ht]
    \centering
    \includegraphics[width=0.47\textwidth]{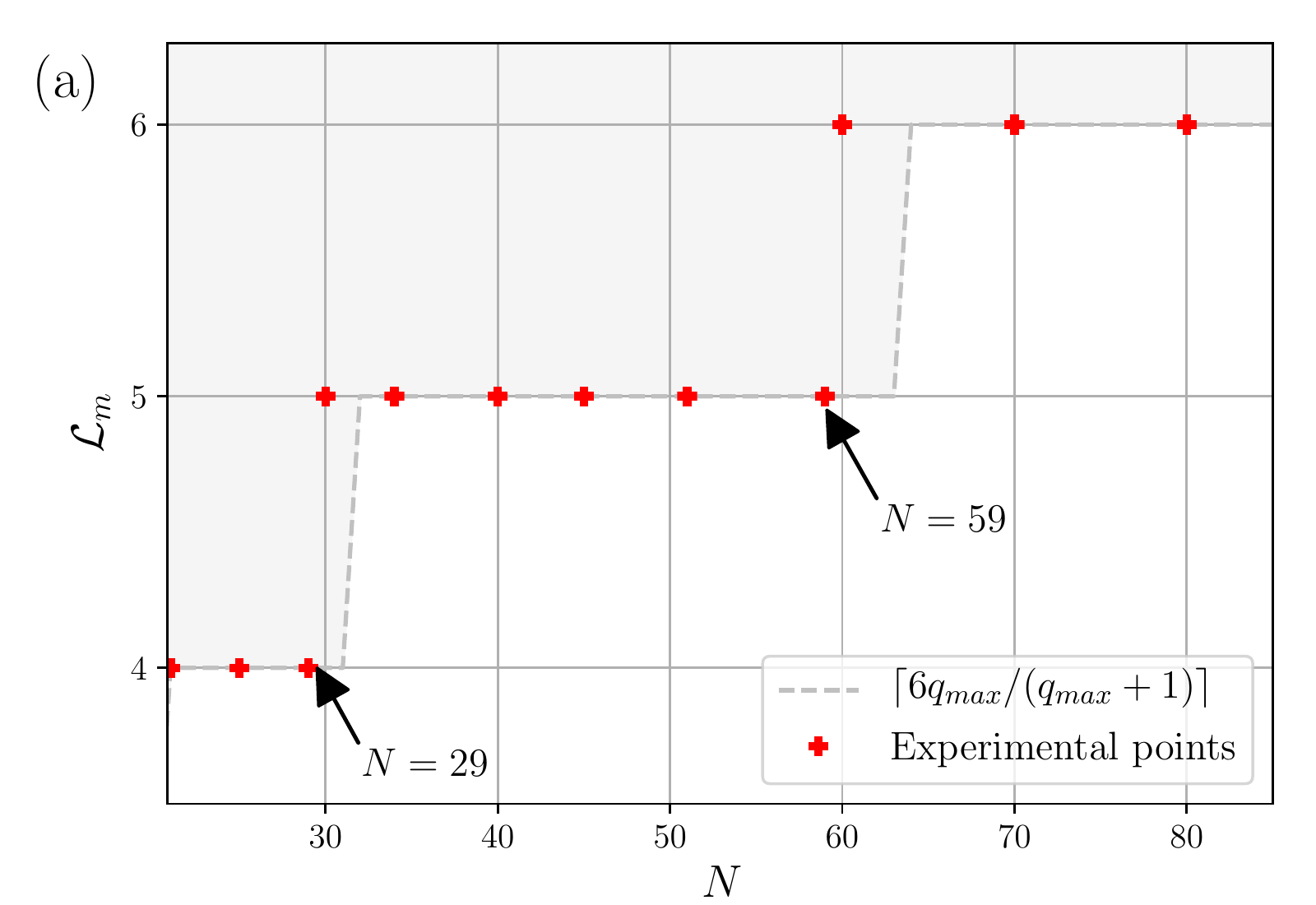}
    \includegraphics[width=0.47\textwidth]{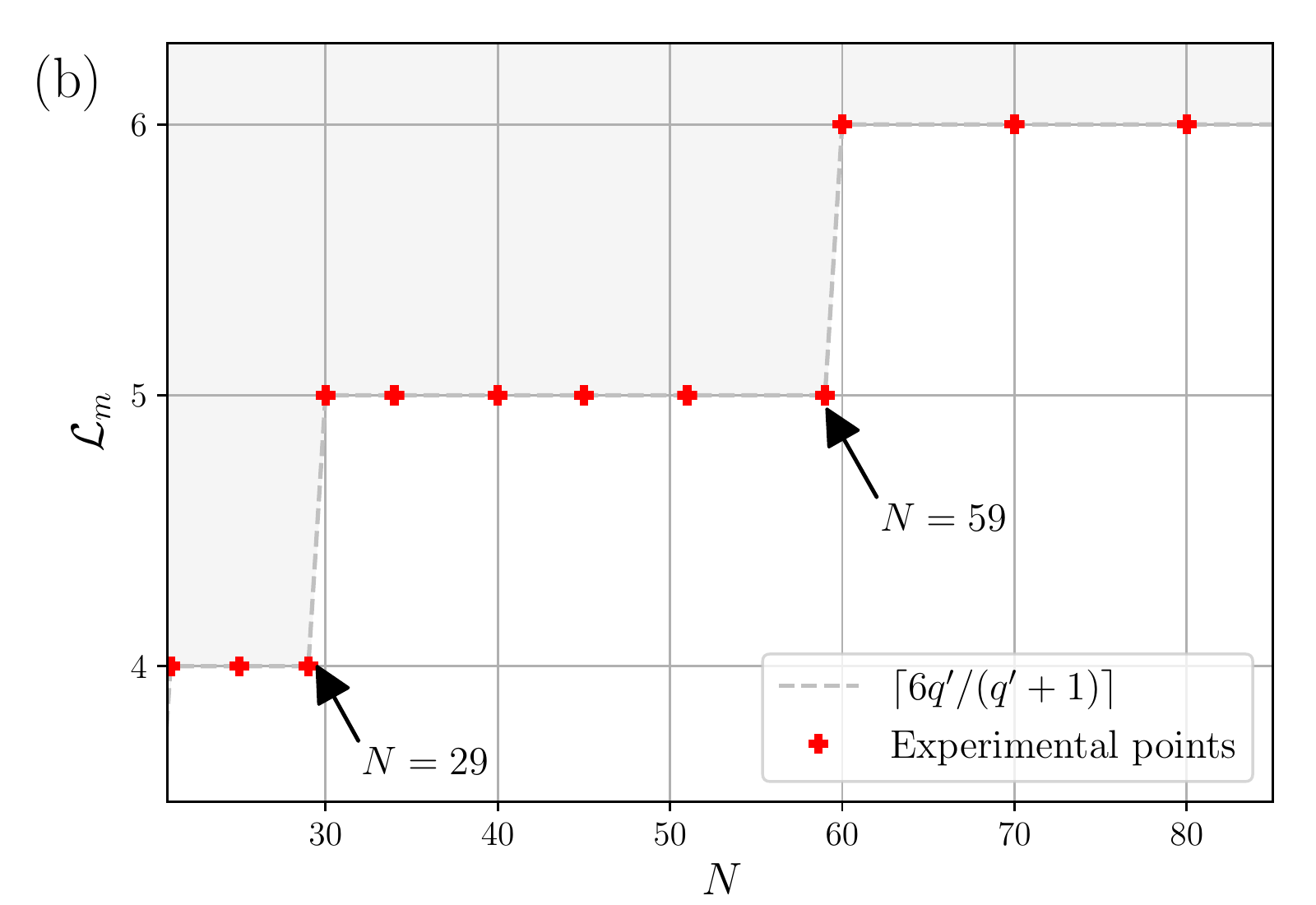} 
    \caption{Minimum number of removed links $\mathcal{L}_m$ for different network sizes $N \in [20,80]$. The red diamonds stand for data obtained via numerical simulations, while the dashed line separating the white and gray regions corresponds to model prediction (Eq. (\ref{eq:Lmin1})). The network parameter controlling the nonlocality is fixed at $R=3$. (a) Comparison between the numerical simulation and the model fed with the parameter $r$ obtained for networks in the thermodynamic limit ($q_{max}(r)$). (b) The model prediction incorporates a correction $\varepsilon$ to account for finite-size networks $q^{\prime} = q_{max}(r-\varepsilon)$.}
    \label{fig:NLinksR3}
\end{figure}

In order to further analyze the ability of Eq. (\ref{eq:Lmin1}) in predicting the results obtained numerically. In Fig. \ref{fig:NLinksR3}, we show a comparison between $\mathcal{L}_m$ obtained numerically and the model for different network sizes. In this analysis, the network coupling vicinity is fixed at $R=3$ while the network size is varied in the interval $N \in [20,80]$. At first, in Fig. \ref{fig:NLinksR3}(a), we observe that $\mathcal{L}_m$ obtained for the model deviates from the numerical simulations (red crosses) for two network sizes, namely $N=30$ and $N=60$. To improve the model performance, we recall that the values of $q_{max}$ fed into the model are obtained from the bifurcation scenario of the Kuramoto system in the thermodynamic limit, see Table \ref{tab:q_and_r}. When compared with finite networks, the bifurcations in this limit usually occur for lower values of $r$. To overcome such an issue, we consider a correction in the parameter $r$ in which its numerical value is decreased by an amount $\varepsilon$ accounting for deviations from the thermodynamic limit.

With this, the largest winding numbers $q_{max}=q_{max}(r)$ in Eq. (\ref{eq:Lmin1}) is shifted in accordance with $\varepsilon$ as $q_{max}(r) \rightarrow q^{\prime} \equiv q_{max}(r-\varepsilon)$. We remark that, as the network size increases, the value of the parameter $r$ obtained here for finite networks approach the values of $r_c$ shown in Table \ref{tab:q_and_r}. As a consequence, the correction $\varepsilon$ depends on $N$. We account for this matter by considering different values of $\varepsilon$ for two different ranges of $N$: 
\begin{equation}
    \varepsilon = \left\{\begin{array}{ll}
        0.0064, & \text{for } N\leq 45\\
        0.0032, & \text{otherwise }\\
        \end{array} \right.
\end{equation}
In Fig. \ref{fig:NLinksR3}(b), we analyze the model performance considering the correction $\varepsilon$ in the parameter $r$. Now, we observe that the minimum number of removed links $\mathcal{L}_m$ for the ISS be globally stable obtained for the model is in good agreement with the numerical data. The general trend in this figure indicates the growth of $\mathcal{L}_m$ for increasing $N$, i.e., it is necessary to remove more links as the network gets less dense. This tendency goes up to a limit at $N\approx60$ in which $\mathcal{L}_m$ equals the number of links $\mathcal{L}$ necessary to completely open the ring. From $N\approx60$ on, the systematic removal of $\mathcal{L}$ links aiming at the opening of the ring topology assures the convergence to the ISS. This is also expressed in our model Eq. (\ref{eq:Lmin1}): for high values of $q_{max}$, $\mathcal{L}_m \approx \mathcal{L}$.

In this subsection, we have verified how a systematic removal of links can drive nonlocally coupled Kuramoto networks towards the ISS. Despite the difficulties of analytically treating such a case with $R>1$, the agreement with simulations of the model based on locally coupled networks suggests the validity for $R>1$ of the mechanisms analytically developed for $R=1$ in subsections \ref{sub:linear_stability} and \ref{sub:bif_analysis}.  

\section{Conclusions}
\label{sec:conclusion}

In summary, we report that the in-phase synchronous state (ISS), sharing the state-space of sparse networks with other solutions, can be globally stabilized by making the network even more sparse. Specifically, for Kuramoto oscillators regularly coupled in a ring topology, we demonstrate that the removal of network links, aiming the opening of ring topology, destabilizes the network $q$-states coexisting with the ISS. We call this phenomenon sparsity-driven synchronization. This claim is supported both numerically and analytically for networks of Kuramoto oscillators. In the numerical approach, we observe the onset of the sparsity-driven synchronization in nonlocally coupled networks. The robustness of the phenomenon is verified against different numbers of coupled nearest neighbors in the network and different sets of initial conditions. Following these results, we have illustrated the employment of this mechanism to drive complete synchronized behavior in the network. Such a control scheme consists in removing network links to make the ring topology open for a time interval. After the system's trajectories reach the globally stable ISS, the original topology is safely restored without removing the trajectories from the sync basin. 

In the analytical approach, we perform the linear stability analysis in the system after the link removal procedure, and we verify that the ISS is the sole network stable state. To complement the analytical study, in the same local network, we introduce a weight to the link intensity making it possible to explore the bifurcation sequences involved in the global stabilization of the ISS. Following this definition, we obtain a low-dimensional description for the network in which the weight is a control parameter. With this, we show that all $q$-twisted states coexisting with the ISS are destroyed in a sequence of saddle-node bifurcations. Furthermore, for nonlocally coupled networks, we investigate the number of links that should be removed in order to ensure global stability to the ISS without the necessity of completely opening the ring. In this context, based on the analytical findings for local networks, we developed a simple model that, given the network parameters, can predict the minimum of removed links that guarantees global stability to the ISS. Finally, the sparsity-driven synchronization is a phenomenon easy to be induced with few interventions in the network, able to rescue the ISS, and, additionally,  also applicable in networks with high connectivity.

\begin{acknowledgments}
  E.S.M. and A.Z. acknowledge support by the Deutsche Forschungsgemeinschaft (DFG, German Research Foundation) - Projektnummer - 163436311-SFB 910. R.O.M.T acknowledge support by S\~ao Paulo Research Foundation (FAPESP, Proc. 2015/50122-0). 
\end{acknowledgments}

\section*{Data Availability}
The data that support the findings of this study are available from the authors upon reasonable request.


%

\end{document}